# HPC-Accelerated Simulation and Calibration for Silicon Quantum Dots

Dhilan Nag
University of Texas, Austin
dhilannag@utexas.edu

Suhun Kim
University of Texas, Austin
suhunkim@utexas.edu

Cole Johnson
Rochester Institute of Technology
cdj5492@g.rit.edu

Collin Sumrell
University of Oklahoma
sumrell@ou.edu

*Abstract*—Quantum computers (QCs) have the potential to solve critical problems significantly faster than today's most advanced supercomputers. One major challenge in realizing this technology is designing robust electrostatic pulses to realize unitaries on qubits. Current practice when calibrating unitaries involves recursive experimentation to find the highest-fidelity pulses. To accelerate this process for experimentalists, we implement Qalibrate, a fast, JAX-enabled simulator that generates pulses given target unitaries. Specifically, we generate a propagator that models the time evolution of three-electron spin qubits and integrate our gradient-based optimizer to generate the pulses. The simulation involves solving the Lindblad master equation, which we parallelize by employing an approximation of the time evolution called the Magnus expansion. Qalibrate shows up to a 34x speedup compared to an existing ODE simulator, making progress towards generating robust pulses for n-qubit systems.

## I. Introduction

Quantum computers (QCs) have the potential to solve critical problems significantly faster than today's most advanced supercomputers—a milestone known as quantum advantage. One major challenge in realizing this emerging technology is precisely controlling the hardware to execute useful algorithms. Quantum algorithms for advanced applications, such as chemical and materials science simulations, require on the order of millions to billions of quantum gates to operate [23]. With current quantum computers in the NISQ (noisy intermediate scale quantum) era [20], the largest available systems can only execute in the order of thousands of quantum gates, far from achieving quantum advantage for these difficult problems.

In working to design QCs that can support longer circuits, quantum gates must be designed that have the highest possible fidelity. Fidelity is a measure of how accurately a physically implemented gate executes a unitary gate operation. To maximize the fidelity of a quantum gate, each qubit in a QC must be characterized and calibrated. Characterization is a process of measuring a qubit's fundamental properties, such as the resonance frequency, relaxation time, and dephasing time. Calibration is the process of using those properties to optimize fidelity, including the generation of a customized pulse sequence for specific quantum gates. Pulse calibration directly impacts the error rates of a quantum gate, and thus, the scalability of a device is limited by the accuracy of calibration.

Most quantum algorithms are designed to work in an ideal world in which quantum gates are perfectly accurate. In the real world, however, gate infidelities are non-zero due to factors including imperfect hardware and environmental noise. Thus, calibration must account for external factors in optimizing pulses to execute a specific quantum gate. Previous work in the field has integrated optimization algorithms such as Bayesian optimization [21] and gradient-based optimization [8], [19] to optimize pulse sequences. There are existing frameworks for realizing these techniques, and much room is still available for practical speedup in important parts of any optimizer.

Gradient-based optimization of pulse sequences has been applied to direct simulation methods [8]. However, forward simulation of the Lindblad master equation is necessarily serial: each subsequent time step must be computed using the state and derivatives at the previous time step. By solving the master equation in the superoperator form, the propagators at different slices of time can be calculated in parallel. Therefore, we propose a differentiable simulator of the Lindblad dynamics that readily parallelizes to multicore systems.

## II. Background

In classical computing, algorithms employ classical logic gates (like AND, OR, and NOT) that manipulate bits of information (0s and 1s). Analogously, quantum computation employs quantum gates, which are operations that manipulate quantum bits, known as qubits. Unlike a classical bit that may only be in either a 0 or a 1 at any time, a qubit may exist in a continuum of both states simultaneously, a phenomenon known as superposition. This is often represented in the form $|\psi\rangle = \alpha |0\rangle + \beta |1\rangle$, where $\alpha$ and $\beta$ represent the the square root of the probability that the qubit is in the 0 or 1 state, respectively, so that $|\alpha|^2 + |\beta|^2 = 1$. Computationally, it is efficient to define the basis states as state vectors $|0\rangle = \begin{bmatrix} 1 \\ 0 \end{bmatrix}$ and $|1\rangle = \begin{bmatrix} 0 \\ 1 \end{bmatrix}$ so that $|\psi\rangle = \begin{bmatrix} \alpha \\ \beta \end{bmatrix}$. Quantum gates can then be represented by unitary matrices that operate on this state vector.

## A. Hardware

Information is fundamentally physical. To run a quantum algorithm in the real world, a qubit must be represented by a physical system. Any quantum entity with two distinct states can be used to represent a qubit. The most promising forms of qubit that have emerged in recent developments include superconducting [15], electron spin [17], photonic [14], trapped ion [3], and neutral atom qubits [11]. Superconducting qubits use distinct non-linear energy levels as quantum states, whereas spin qubits use the distinct spins of an electron.

SLEDGE (Single-Layer Etch Defined Gate Electrode) devices [9] are a specific kind of spin NISQ device. In this device, three trapped and isolated electrons, called dots, are used to collectively represent the quantum states of a qubit. Physically realizing gates on such a qubit is done through dot-dot exchange interactions, often achieved by varying barrier potentials between electron wells, resulting in Coulombic interactions between electrons that can yield entangled states. The Hubbard Hamiltonian models the dynamics of this exchange interaction of any number of quantum dots [10].

$$\hat{H} = \sum_{i=1}^{N} \left[ \frac{\tilde{U}}{2} \hat{n}_i(\hat{n}_i - 1) + V_i \hat{n}_i \right] \\ + \sum_{\langle i,j \rangle} U_C \hat{n}_i \hat{n}_j \\ + \sum_{\langle i,j \rangle} \sum_{\sigma=\uparrow,\downarrow} t_{ij} \left( \hat{c}_{i\sigma}^\dagger \hat{c}_{j\sigma} + \hat{c}_{j\sigma}^\dagger \hat{c}_{i\sigma} \right) \quad (1)$$

In this model, $\hat{c}_{i,\sigma}^\dagger$ is the creation operator and $\hat{c}_{i,\sigma}$ is the destruction operator for an electron with spin $\sigma$ at quantum dot i. The number operator $\hat{n}$ is defined by $\sum_\sigma \hat{c}_{i,\sigma}^\dagger \hat{c}_{i,\sigma}$, which yields the number of electrons in a given quantum dot. $\tilde{U}$ describes the penalty for doubly occupied dots, V is the chemical potential that favors or does not favor a presence of an electron in the quantum dot, $U_C$ is the energy penalty for electrons in neighboring quantum dots, and $t_{ij}$ is the hopping term that describes electrons tunneling between adjacent quantum dots. Hence, the Hubbard model is appropriate in extensive modeling of the quantum dot architecture where each dot has a four-dimensional Hilbert space: $\{|\ \rangle, |\uparrow\rangle, |\downarrow\rangle, |\uparrow\downarrow\rangle\}$. In other words, each quantum dot can have no electron, a single spin-up or spin-down electron, or both spin-up and spin-down electrons [22]. States like $\{|\uparrow\uparrow\rangle, |\downarrow\downarrow\rangle\}$ are excluded from the Hilbert space due to the Pauli exclusion principle. When considering tunable variables in the real world to generate pulse sequences, it is natural to adjust the $t_{ij}$ parameters associated with the inter quantum dot hopping term.

In the case of singlet-triplet qubit [5], the logical state $|0\rangle$ is encoded with the physical singlet state defined by the ket $|S\rangle = \frac{1}{\sqrt{2}}(|\uparrow\downarrow\rangle - |\downarrow\uparrow\rangle)$ and the logical $|1\rangle$ is encoded with the triplet state $|T_0\rangle = \frac{1}{\sqrt{2}}(|\uparrow\downarrow\rangle + |\downarrow\uparrow\rangle)$. For the singlet-triplet qubit, however, the degeneracy between the three triplet states that share the same energy needs to be broken by an external magnetic field gradient. Thus, in addition to electrical control of the exchange interactions, magnetic field control is required to manipulate quantum states.

Three spin qubits provide an advantage that all state manipulation is done via the exchange interactions [4] and therefore all control is electrical in nature. The logical qubit states are encoded in the decoherence-free subspace [16], meaning that the global magnetic noise that affects all three electrons in the same way does not affect the total quantum state of the system. When it is set that each quantum dot contains one up or down spin electron, the system becomes an eight-dimensional Hilbert space. The eight-dimensional Hilbert space can be decomposed into a four-dimensional $S = \frac{3}{2}$ space and degenerate $S = \frac{1}{2}$ doublet space. Once the degeneracy is lifted for the doublet space with a uniform magnetic field, each $S = \frac{1}{2}$ doublet provides a two-level system in which the logical qubits can be encoded. Thus, a logical qubit can be encoded as follows:

$$|0\rangle = \frac{1}{\sqrt{2}} (|\uparrow\uparrow\downarrow\rangle - |\downarrow\uparrow\uparrow\rangle) \quad (2)$$

$$|1\rangle = \frac{1}{\sqrt{6}} (2|\uparrow\downarrow\uparrow\rangle - |\uparrow\uparrow\downarrow\rangle - |\downarrow\uparrow\uparrow\rangle) \quad (3)$$

After defining the logical basis in the physical space, we can tune the exchange interaction parameters to adjust the quantum states.

## B. Quantum Dynamics

In the previous section, we discussed the relevant physics associated with the spin qubit. The Hamiltonian is essential in finding the pulse sequence of a quantum gate because the Hamiltonian can fully describe the time evolution of a quantum state in a closed system. The Schrödinger equation defines this time-evolution relation:

$$i\hbar \frac{d}{dt} |\psi(t)\rangle = \hat{H} |\psi(t)\rangle \quad (4)$$

With an appropriate Hamiltonian, by solving the Schrödinger equation, we can find the final quantum state with an initial state. The Schrödinger equation, however, describes the unitary evolution of pure states. In many cases of quantum system control, mixed states are involved, meaning that there is more than one state, each with different probabilities, instead of one pure state undergoing time evolution. Thus, using density matrices instead of pure states is advantageous because it allows us to represent an ensemble of quantum states. A density matrix can be derived from a pure state where $\rho = |\psi\rangle \langle \psi|$. The off-diagonal elements of a density matrix represent coherence, or the superposition of different basis states.

The time evolution of a density matrix can be represented by the von Neumann equation, given by:

$$i\hbar \frac{d\rho(t)}{dt} = [\hat{H}, \rho(t)] \quad (5)$$

The von Neumann equation states that the commutator of the Hamiltonian and the density matrix is associated with the change of the density matrix with respect to time. In other words, if the density matrix and the Hamiltonian commute, or the density matrix is composed of the energy eigenstates of the Hamiltonian, then such a density matrix will not change over time. However, in the case that the density matrix and the Hamiltonian do not commute, the density matrix undergoes time evolution, typically involving oscillations or decoherence between energy eigenstates.

The von Neumann equation applies to the full density matrix describing both the system of interest and the environment's joint quantum state. Typically, approximations are made that model the action of the environment on the system of interest via the action of Markovian jump operators. Under this approximation, the density matrix of the system has dynamics governed by the Lindblad master equation [18]:

$$\frac{d\rho}{dt} = -\frac{i}{\hbar}[H, \rho] + \sum_k \left( L_k \rho L_k^\dagger - \frac{1}{2}\{L_k^\dagger L_k, \rho\} \right) \quad (6)$$

The first term of the Lindblad master equation describes the action of the time-dependent Hamiltonian on the density matrix, while the $L$ terms are the jump operators or the collapse operators that represent the dissipative dynamics caused by the environment. The jump operators are crucial because they make the system evolution non-unitary and model the spontaneous dynamics, such as relaxation and dephasing. For instance, jump operators can be created from the $T_1$ (time for an excited qubit to relax) and $T_2$ (time for a qubit to lose phase information) as the following: $L_1 = \frac{1}{\sqrt{T_1}}a$, $L_2 = \frac{1}{\sqrt{T_2}}a^\dagger a$ where $a$ and $a^\dagger$ are the lowering and raising operators acting on the system. Hence, by solving the Lindblad master equation, we can solve for the final evolved density matrices considering dissipative dynamics of the system.

C. Experimental Calibration

Calibration methods for obtaining the model dynamics are typically based on experimentation. A range of electrostatic pulse waveforms is applied to qubits, and the corresponding qubit rotations are recorded for experimentalists to reference when executing gates. Recalibration is required at time scales on the order of hours since errors drift across the hardware such that qubits have different responses to pulses over time. The process of testing a range of pulses and recalibrating more frequently than is currently standard has the potential for increased continued fidelity. Qalibrate reduces the experimental overheads by generating waveforms given the desired rotation gates.

D. Forward ODE Solvers

Existing pulse generators such as Quandary [8] simulate hardware by solving the Schrödinger equation or Lindbladian forward in time. Quandary approximates the propagator's behavior by evolving three specially chosen states, and generalizes the time evolution for all states. Such generalization can avoid increasing the dimensions of the propagator because the quantum state can evolve in the Hilbert space and not in the Liouville space. However, explicitly computing the superpropagator yields an exact time evolution of all quantum states in open dynamics.

III. Methodology

A. Propagators

General transformations of density matrices that obey the axioms of quantum mechanics are described by completely positive and trace-preserving maps. The Lindblad equation, together with an initial density matrix, defines an initial value problem that can be solved numerically via ODE solvers. Therefore, this initial value problem must be solved in serial: the state, described by the density matrix at each time, $\rho(t)$, must be propagated to the next moment in time by an equation that depends on its current state.

To describe the map which takes any initial density matrix into its final state according to the time-dependent Hamiltonian and jump operators is possible using the superoperator formalism [18]. In this case, the density matrix is flattened or 'vectorized', and typically denoted $|\rho\rangle\rangle$, and its dynamics are given by

$$\frac{d}{dt}|\rho(t)\rangle\rangle = \mathcal{L}(t)|\rho(t)\rangle\rangle, \quad (7)$$

where the Liouvillian superoperator $\mathcal{L}(t)$ is given by the Hamiltonian and jump terms via

$$\begin{aligned}\mathcal{L}(t) = &-i\big(I \otimes H(t) - H(t)^\mathsf{T} \otimes I\big) \\ &+ \sum_k \big[ L_k(t)^* \otimes L_k(t) - \tfrac{1}{2} I \otimes L_k(t)^\dagger L_k(t) \\ &- \tfrac{1}{2}\big(L_k(t)^\dagger L_k(t)\big)^\mathsf{T} \otimes I \big]. \end{aligned} \quad (8)$$

Formally, this solution can be solved by introducing an operator $\mathcal{U}(t)$, which solves

$$\frac{d}{dt}\mathcal{U}(t) = \mathcal{L}(t)\mathcal{U}(t) \quad (9)$$

together with the constraint of continuity,

$$\mathcal{U}(0) = \mathbf{1}. \quad (10)$$

This ODE has a formal solution given by

$$\mathcal{U}(t) = \mathcal{T} \exp\left( \int_0^t \mathcal{L}(t')dt' \right) \quad (11)$$

where $\mathcal{T}$ denotes a causal time-ordering of all the operators in the integral. This time-ordering is necessary because the Liouville superoperator at different times will, in general, not commute:

$$[\mathcal{L}(t_a), \mathcal{L}(t_b)] \neq 0. \quad (12)$$

## B. Magnus Expansion

The Magnus expansion has been applied to numerically solving large closed quantum systems previously [2]. In that case, the Magnus expansion was used to solve the propagator for the Schrödinger equation, and is therefore limited to describing closed quantum systems.

To parallelize computing the propagator, we employ the Magnus expansion. By describing the Lindblad equation in the superoperator form, it satisfies a first-order homogeneous linear differential equation, and we are able to apply the Magnus expansion to approximate its solution.

This method splits the time-dependence into slices, and averages the Liouvillian superoperator over the time-dependence within each slice [1], which are solved simultaneously using the JAX library to produce the propagator for each chunk:

$$\overline{\mathcal{L}}_{t_{n+1},t_n} = \int_{t_n}^{t_{n+1}} \mathcal{L}(t)\,dt \quad (13)$$

$$U_{t_{n+1},t_n} = \mathcal{T}\exp\left(\overline{\mathcal{L}}_{t_{n+1},t_n}\right) \quad (14)$$

The separate propagators for each chunk can then be reduced to a single propagator through successive matrix multiplications:

$$U_{t_n,t_0} = U_{t_n,t_{n-1}} U_{t_{n-1},t_{n-2}} \cdots U_{t_1,t_0} \quad (15)$$

Matrix multiplication is associative, so this multiplication chain can be efficiently computed using a tree reduction provided by the JAX library.

## C. Gradient-Based Optimization

We start with an initial set of parameters to optimize, which is the coefficient of the control Hamiltonians that represent the barrier gate voltages. Then, the Magnus superpropagator is computed using an initial coefficient matrix of dimensions $(M, N-1)$, where $M$ is the number of Magnus slices and $N$ is the number of quantum dots. We define our loss function as the mean squared error difference between the computed propagator and the target unitary, a matrix form of an intended quantum gate. However, the dimensions of the propagator and the target unitary are inherently different. The propagator exists in the Liouville space, whereas the target unitary exists in the logical space. Hence, a projection is necessary to compare the two matrices in equal dimensions [7].

In the case of three quantum dots with well-defined logical encoding, we can generate a projection matrix consisting of the two basis states that represent the logical qubit encoding in the physical 64-dimensional Hilbert space (each quantum dot has a 4-dimensional Hilbert space). Thus, enumerating all possible states of the total Hilbert space, the basis state vector $\phi_0$ representing $|0\rangle = \frac{1}{\sqrt{2}}(|\uparrow\uparrow\downarrow\rangle - |\downarrow\uparrow\uparrow\rangle)$, would have coefficients $\frac{1}{\sqrt{2}}$ and $-\frac{1}{\sqrt{2}}$ in indices 21 and 33. We do the same to get a basis vector for $|1\rangle$. Liouville space works with flattened density matrices, so we further create a density matrix of the basis state vector by taking the outer product. $\rho_0 = |\phi_0\rangle\langle\phi_0|$, $\rho_1 = |\phi_1\rangle\langle\phi_1|$. The resulting projection matrix $P$ is a matrix composed of two column vectors, which are flattened density matrices $\rho_{f0}$, $\rho_{f1}$. The loss function computes the mean squared error between the target unitary and the propagator projected down to the logical space, and

$$\Pi = P^\dagger U P \quad (16)$$

where $\Pi$ is the projection and the U is the propagator. Thus, the loss function computes the mean squared error of the $\Pi$ and the target unitary by calculating the Frobenius norm of the difference between the two matrices defined by:

$$\|A\|_F = \sqrt{\operatorname{trace}(A^\top A)} \quad (17)$$

where $A$ is the difference matrix. We then use the gradient-based optimization algorithm Adam [13], implemented in the Optax library, to update the coefficients of the control Hamiltonian. The updated pulses are used to compute the propagator, and the optimization loop proceeds until the losses are minimized.

## IV. Evaluation

Qalibrate's Magnus expansion constructs the propagator faster than the Dynamiqs [6] mepropagator function because of its parallelization. Figure 1 shows the runtime for propagator generation for up to 3 quantum dots, which represent one qubit on a SLEDGE device. Qalibrate's method takes 1 minute and 15 seconds to run 3 quantum dots while Dynamiqs takes 43 minutes, showing a 34x speedup. The performances are expected to diverge further as larger numbers of dots are tested since Dynamiqs continues to solve the Lindbladian forward in time.

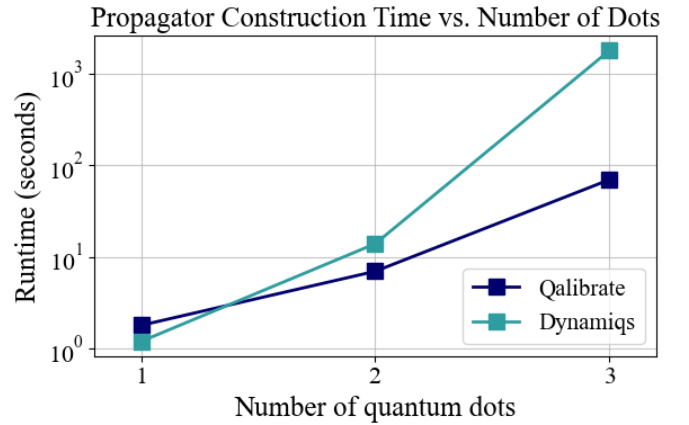

Fig. 1. Propagator construction time vs. number of quantum dots.

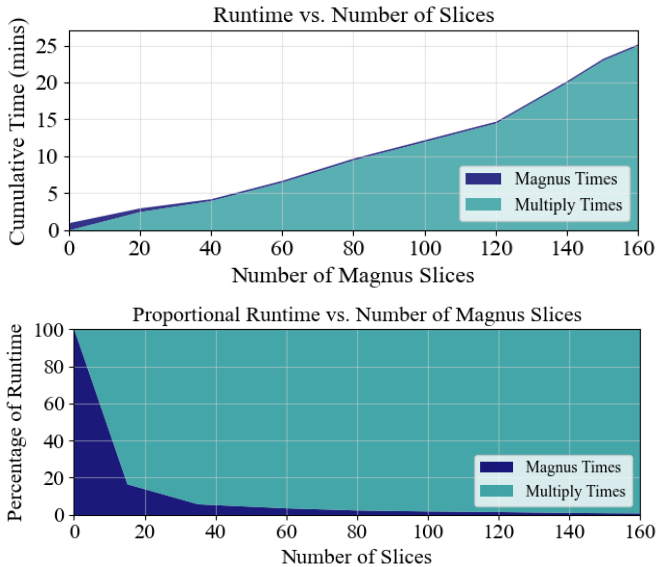

Fig. 2. Propagator construction time increases with the number of Magnus slices. Each slice is computed on a single-threaded AMD EPYC 9634 core.

Qalibrate's propagator generation consists of two steps: Magnus expansion and multiplication. In the Magnus expansion step, the Lindbladian is sliced at discrete time intervals such that all slices are computed as shards in parallel on separate cores. Then, all intermediate propagators are reduced by multiplication to generate a final propagator. Figure 2 shows that as we increase the number of Magnus slices, matrix multiplication dominates the time complexity, and computing the Magnus superpropagators only takes a small fraction of total time.

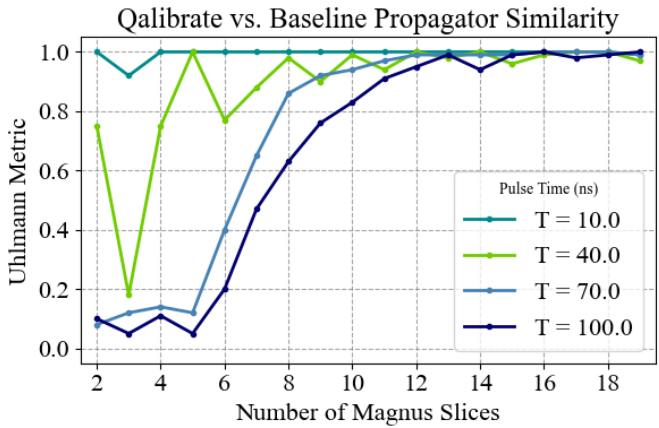

Fig. 3. The Magnus expansion and Dynamiqs propagators are applied to the same density matrices, and the similarity between the matrices is computed using the Uhlmann metric. The Magnus output converges to the Dynamiqs output as the number of Magnus slices increases.

Figure 3 shows how we test the accuracy of our implementation of Magnus expansion by comparing it to the time evolution of a state in QuTiP [12]. Starting with an initial random density matrix, we evolve the density matrix through the propagators generated by QuTiP and the Magnus approach and compare the final density matrices. We employ the Hubbard Hamiltonian for both time evolutions and use QuTiP's mesolve function to get its final state. As the number of Magnus slices increases, the final density matrix produced by applying the Magnus expansion's propagator converges to QuTiP's resulting density matrix. Increasing the number of Magnus slices results in a more accurate final state because the time evolution occurs at a smaller step size. Moreover, for a shorter length time evolution, the Magnus expansion converges even at a very small number of Magnus slices. However, as we increase the total length of the time evolution, more Magnus slices are required to capture the evolution accurately. This relationship is analogous to the accuracy of a Riemann sum and the size of its component rectangles.

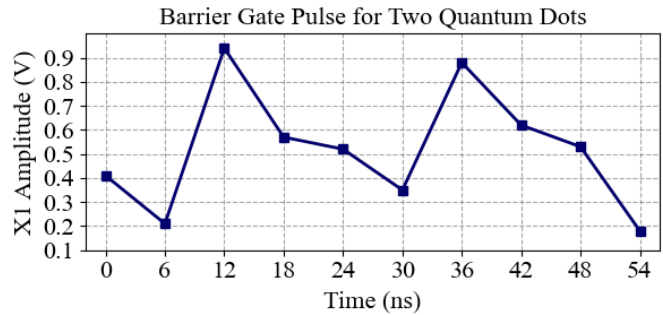

Fig. 4. Example pulse sequence for two quantum dots.

Although the current design does not fully integrate the gradient optimizer into the propagator generator, Qalibrate was able to generate a sample pulse sequence for two quantum dots output from the gradient optimizer as shown in Figure 4. The pulse sequence does not resemble any quantum gate because, at a minimum, a one-qubit gate would require a simulation of three quantum dots. Instead, it shows that we were able to generate an output by integrating the propagator generator and the gradient optimizer. The main issues with the optimizer were a memory issue due to gradient computation requiring intermediate states to be saved. Moreover, the propagator is computed inside the loss function, so every mathematical operation dependent on the coefficients of the control Hamiltonian needs to be differentiated. Such issues arose in integrating the gradient optimizer, and we were not able to generate pulse sequences for three or more quantum dots.

## V. Discussion

Although Qalibrate shows time-scaling improvements for gate calibration, qubit characterization remains a bottleneck for this solution. Voltage pulses must still be manually applied to qubits to generate a profile of the hardware. Qalibrate ingests the characterization profile to rapidly generate propagators, but building tools to accelerate the characterization process is also necessary. Future work could attempt to mitigate this problem by automating the characterization process to profile the hardware at consistent time intervals. Such a solution would likely require interfacing accelerated software techniques with actuation.

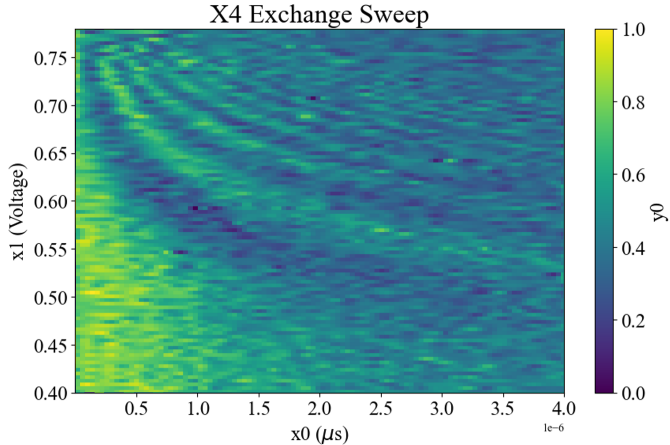

Fig. 5. Sample characterization of SLEDGE device.

The jump operators used in Qalibrate's Hamiltonian are derived from the characterization data shown in Figure 5. The figure shows the normalized waveform response for increasing voltage pulses. The vertical axis denotes the tested voltages, while the horizontal axis denotes the time evolution of the output waveform. The coloration represents the normalized amplitude of the waveform at a given time and voltage. Taking a slice of this heat map at a target voltage generates a dissipating sinusoidal waveform that can be defined using a curve-fitting regression model. Such models enabled us to derive the $T_1$ and $T_2$ times used in the jump operators, but these constants can only be as accurate as the characterization data. This limitation motivates the need for not only faster but also more accurate characterization techniques in future work.

## VI. Conclusion

Qalibrate exhibits promise in solving the problem of generating pulse sequences for a three-electron exchange-only qubit device. The solution was divided into two parts: an accelerated propagator generator that models the open system dynamics via the Lindblad master equation, and a gradient-based optimizer that optimizes the propagator to resemble the desired quantum gate. Qalibrate achieved a significant speedup in generating the propagator compared to the baseline of the Dynamiqs mepropagator function by employing the Magnus expansion and computing the propagators in parallel.

Although the current design does not fully integrate the propagator generator and gradient-based optimizer due to memory limitations and the complexity of differentiating through complex mathematical operations, our framework demonstrates proof-of-concept for two-dot systems and lays a foundation for scaling to three or more dots. Future work will further investigate the temporal and spatial complexity of the gradient optimizer to improve its scalability, and ultimately generate robust pulses for a single qubit gate or even a two-qubit gate.